%% file: briere.tex
\documentclass[12pt]{article}
\usepackage{graphicx}

\def\cmu{Department of Physics\\
Carnegie Mellon University, Pittsburgh, PA 15213 USA}
\def\support{\footnote{Work supported in part by US DOE DE-FG02-91ER40682.}}

\def\Title#1{\begin{center} {\Large #1 } \end{center}}
\def\Author#1{\begin{center}{ \sc #1} \end{center}}
\def\Address#1{\begin{center}{ \it #1} \end{center}}

\newenvironment{Abstract}{\begin{quotation}  }{\end{quotation}}
\newenvironment{Presented}{\begin{quotation} \begin{center} 
             PRESENTED AT\end{center}\bigskip 
      \begin{center}\begin{large}}{\end{large}\end{center} \end{quotation}}
\def\Acknowledgements{\bigskip  \bigskip \begin{center} \begin{large}
             \bf ACKNOWLEDGEMENTS \end{large}\end{center}}

\input econfmacros.tex

\begin{document}
\begin{titlepage}


\vfill
\Title{Highlights from BESIII}
\vfill
\Author{Roy A. Briere\support\\
 {\it for the BESIII Collaboration}}
\Address{\cmu}
\vfill
\begin{Abstract}
We present a selection of recent results from the BESIII collaboration, 
including both charmonium and $D$ meson physics.  
We first discuss the observation of a charged, charmonium-like state, 
the $Z_c(3900)$.  
Conventional charmonium topics include a search for lepton flavor violation, 
studies of $\chi_{cJ} \to \gamma\gamma$ decays, 
and mass and width determinations 
for the $h_c$, $\eta_c(1S)$, and $\eta_c(2S)$.  
We finish with results on the decay constant $f_D$ from $D^+ \to \mu^+ \nu$ 
and on form-factors in $D^0 \to K^- e^+ \nu, \pi^- e^+ \nu$.  
\end{Abstract}
\vfill
\begin{Presented}
Flavor Physics and CP Violation 2013\\
Buzios, Brazil 19-24 May, 2013
\end{Presented}
\vfill
\end{titlepage}
\def\thefootnote{\fnsymbol{footnote}}
\setcounter{footnote}{0}

\section{Introduction}

The BESIII experiment at the BEPCII collider has accumulated the world's 
largest datasets at charm threshold.  
Results discussed here are based on samples of 225 million $J/\psi$ decays, 
106 million $\psi(3686)$ decays, and 2.9 fb$^{-1}$ at the $\psi(3770)$.  
These datasets are approximately 4x, 4x, and 3.5x, respectively, 
compared to the previous best and represent the first iteration of 
our charmonium and $D$ meson programs.  

In addition to the core physics summarized above, BESIII is able to 
do additional physics at other energies, including but not limited 
to $R_{had}$ energy scans and precision $\tau$ mass measurements.  
Indeed, we will start this presentation with our most surprising result, 
based on a fraction of our 2013 data taken at a center-of-mass energy 
corresponding to the $Y(4260)$ state.

\section{\boldmath $Z_c(3900)$: An Exotic Charged State?}

The Belle collaboration recently observed two exotic charged bottomonium-like 
states, the $Z_b(10610)$ and $Z_b(10650)$\cite{Z_b}.  
These states were observed in the $\Upsilon(nS)\,\pi^\pm$ 
and the $h_b(mS)\,\pi^\pm$ mass spectra in the decays 
$\Upsilon(5S) \to \Upsilon(nS)\,\pi^+\pi^-$ and 
$\Upsilon(5S) \to h_b(mS)\,\pi^+\pi^-$ (here, $n = 1,2,3$ and $m=1,2$).  

BESIII recently accumulated a 0.525 fb$^{-1}$ dataset at a center-of-mass 
energy of 4260 MeV.  
While studying the $J/\psi\,\pi^+\pi^-$ final state, a well-known 
decay mode of the $Y(4260)$, a new structure was observed 
in our $J/\psi\,\pi^\pm$ mass spectra\cite{Z_c}.  

In Fig. \ref{fig:Zc} we display both the $J/\psi\,\pi^+$ and $J/\psi\,\pi^-$ 
mass distributions.  We observe two similar peaks in each plot.  
Studies of Monte-Carlo samples show that the lower-mass peak is a reflection 
of the higher-mass peak.  This is due to the fact that the Dalitz plot of 
$M^2(J/\psi\,\pi^+)$ vs. $M^2(J/\psi\,\pi^-)$ is a rather narrow elongated 
band.  
We also note significant structure in the $\pi^+\pi^-$ mass, but find that 
this can be modeled with only a few amplitudes without leading to 
significants changes in the $J/\psi\pi^\pm$ mass distributions as 
compared to phase space; see Fig. \ref{fig:Zc}.  
In addition to occurring in both $J/\psi\,\pi^+$ and $J/\psi\,\pi^-$, 
the $Z_c$ peak also occurs with both $e^+e^-$ and $\mu^+\mu^-$ decays 
of the $J/\psi$, and with both the low-mass and high-mass lobes 
of the $\pi^+\pi^-$ mass distribution.  

{
\begin{figure}
\centering
\includegraphics[width=0.32\textwidth]{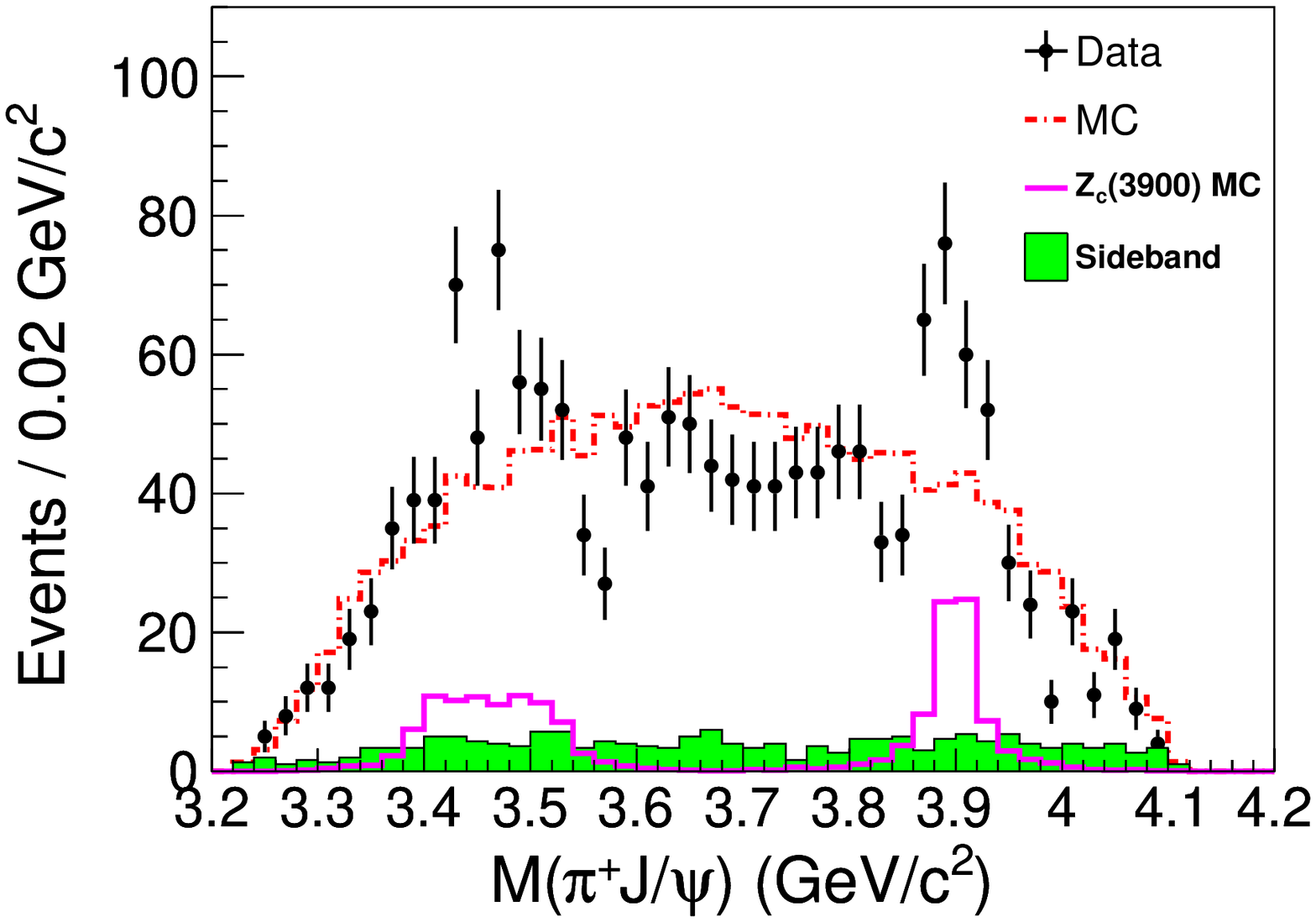}
\includegraphics[width=0.32\textwidth]{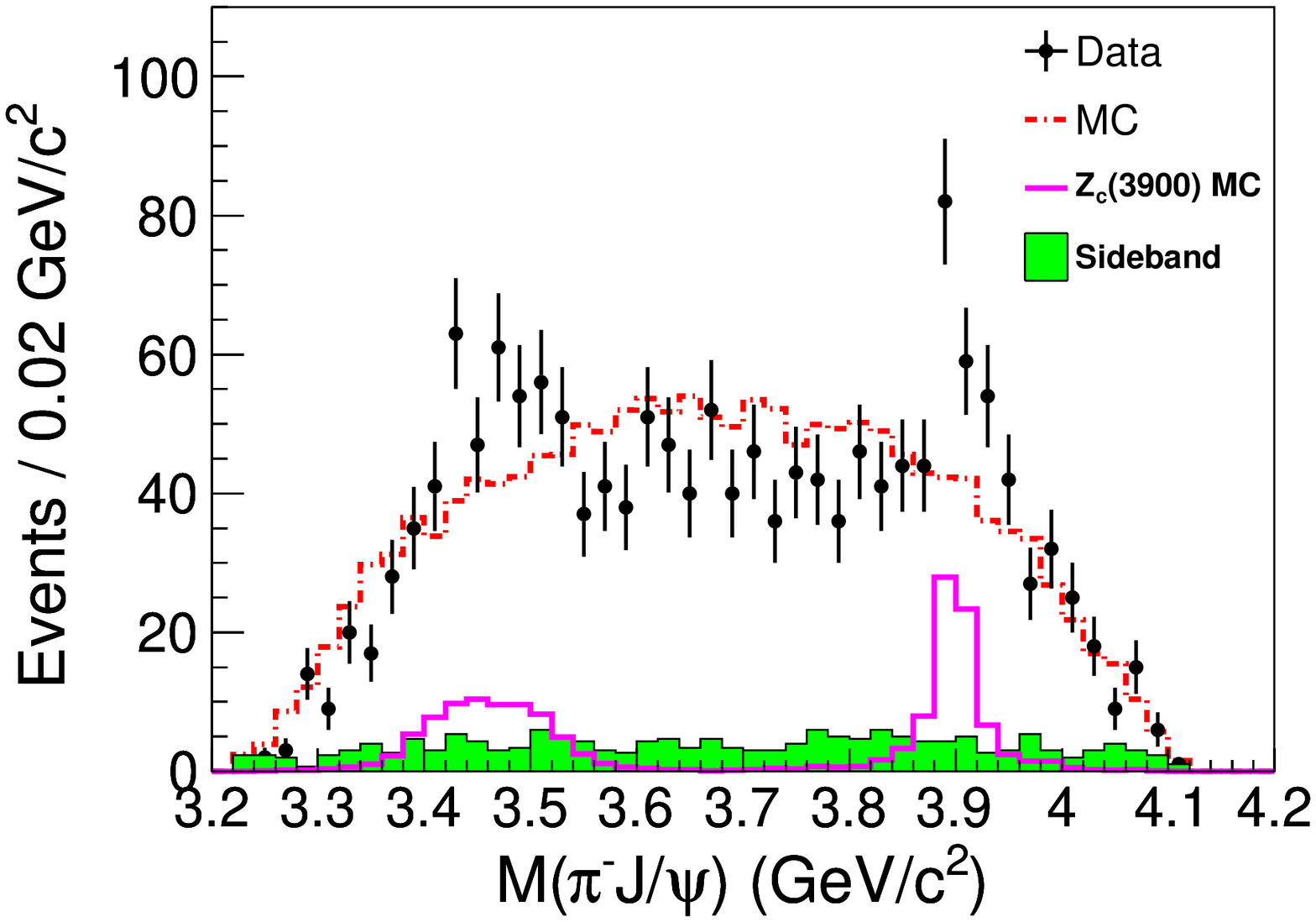}
\includegraphics[width=0.32\textwidth]{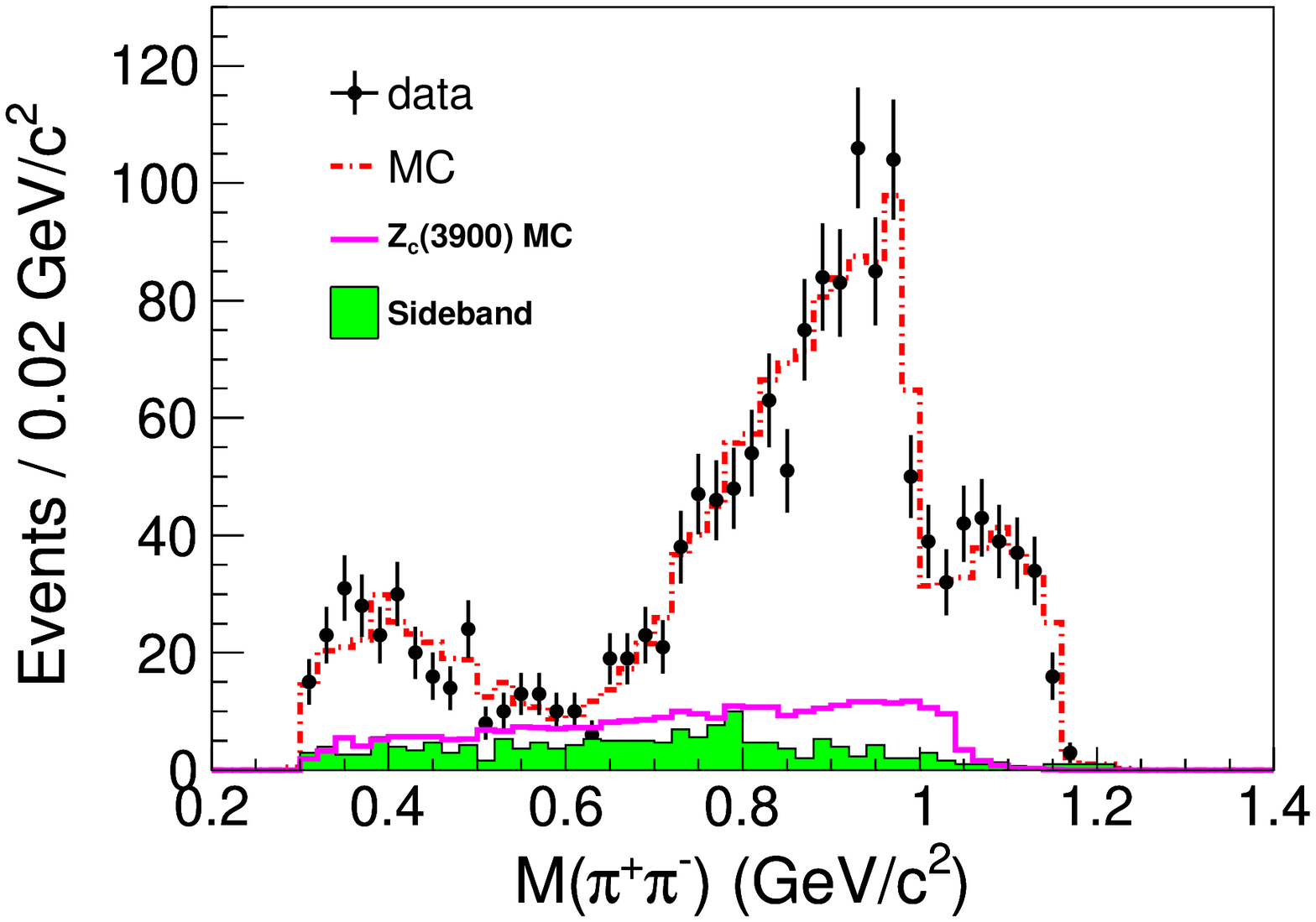}
\caption{Plots of $J/\psi\,\pi^+\pi^-$ events.  
Right: The $J/\psi\,\pi^+$ mass; 
center: the $J/\psi\,\pi^-$ mass; 
left: the $\pi^+\pi^-$ mass.   
In each plot, the points are data, 
the green filled histogram is background estimated from $J/\psi$ sidebands, 
the red dashed line is a Monte-Carlo simulation including 
$f_0(980)$, $\sigma(500)$, and non-resonant terms in the di-pion mass, 
and the magenta line is $Z_c(3900) \pi^\pm$ Monte-Carlo.  }
\label{fig:Zc}
\end{figure}

\begin{figure}
\centering
\includegraphics[width=0.48\textwidth]{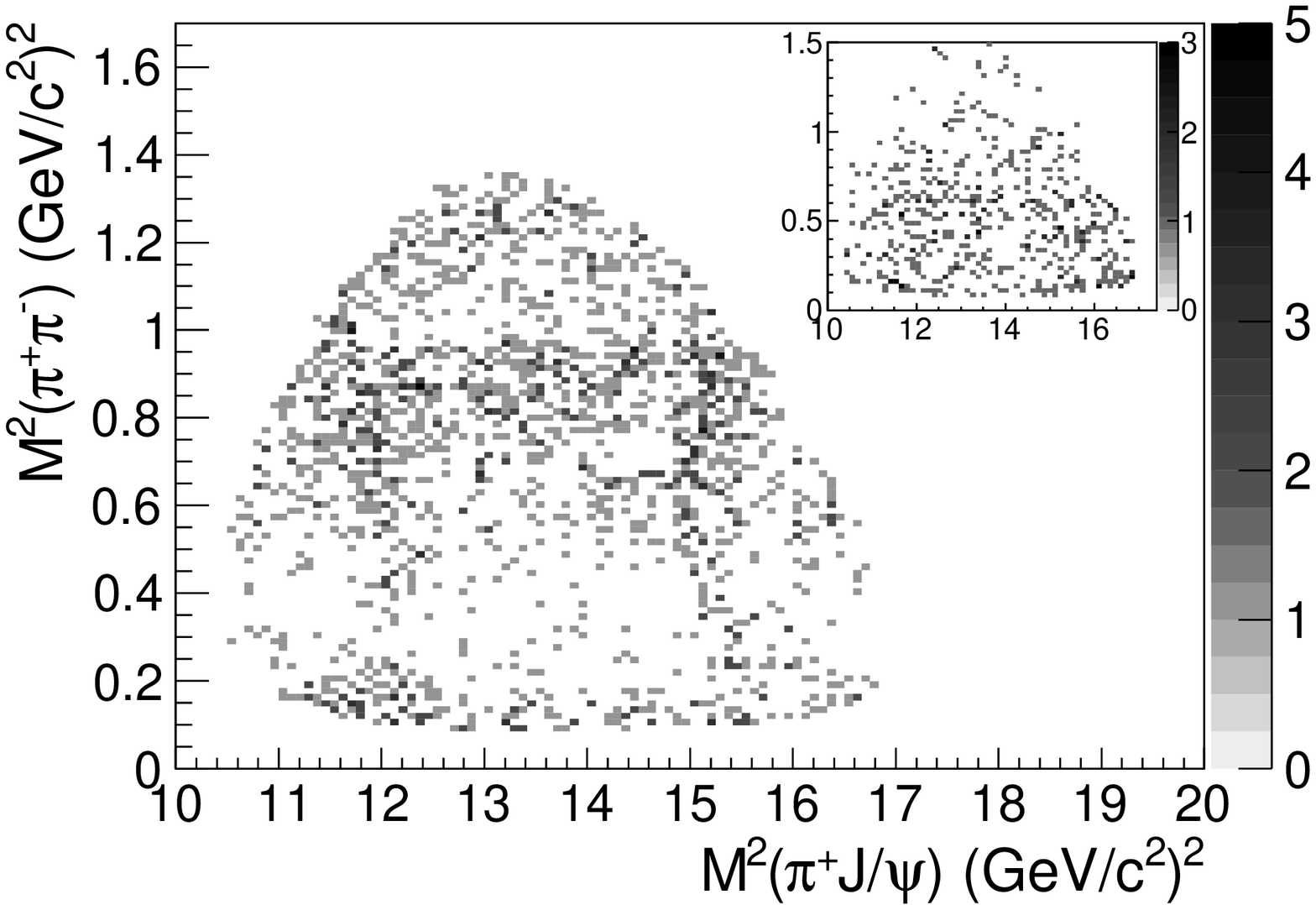}
\includegraphics[width=0.48\textwidth]{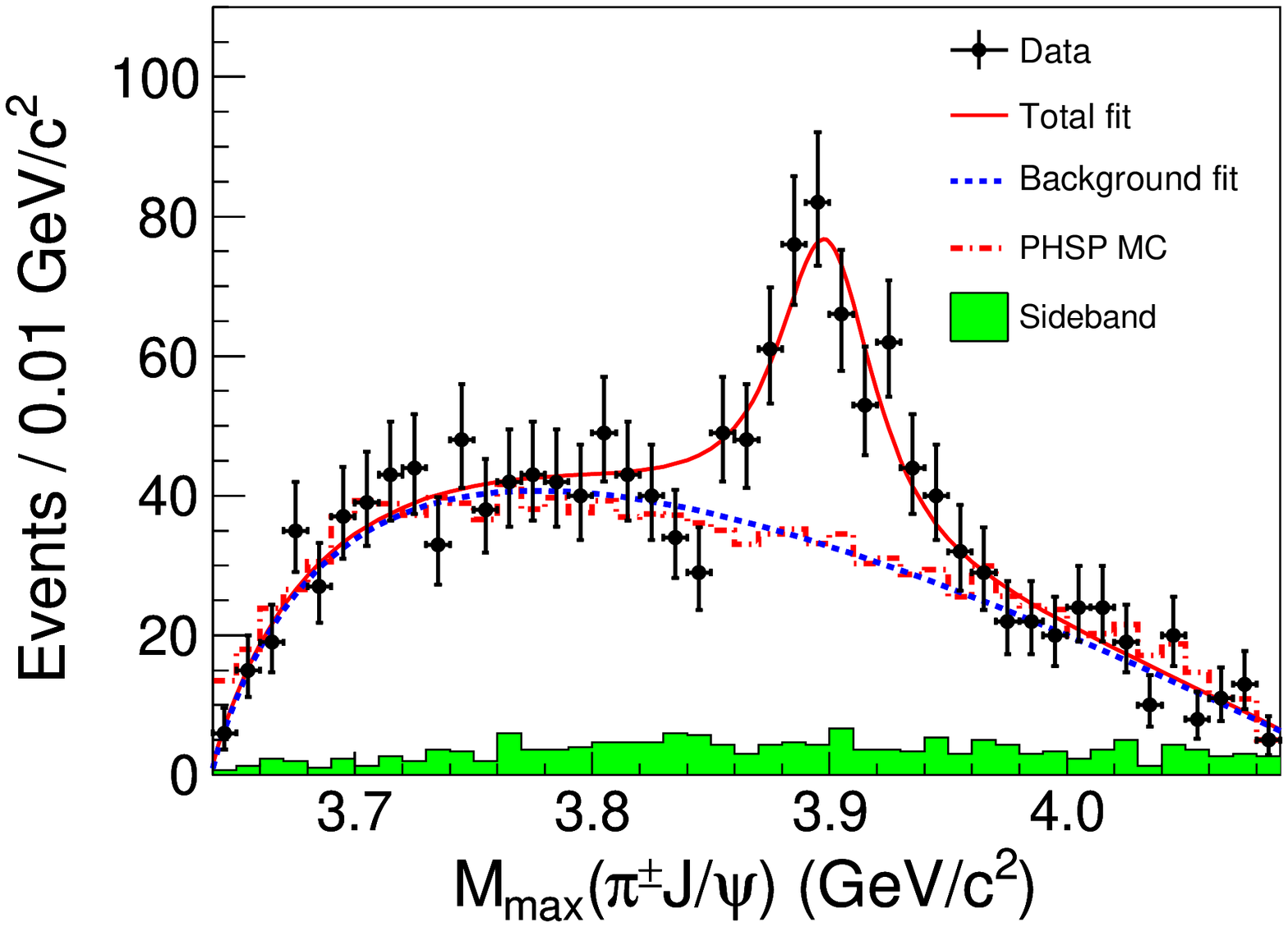}
\caption{Left: Dalitz plot of $J/\psi\,\pi^+\pi^-$ events.  
Note the $Z_c$-related vertical bands and the 
horizontal variations vs. the $\pi\pi$ mass. 
Right: A plot of the larger of the $J/\psi\,\pi^\pm$ masses, 
fit to a $Z_c(3900)$ resonance term and an empirical 
background (only slightly different from phase space).}
\label{fig:Zcfit}
\end{figure}
}

In subsequent analysis, we make a single $J/\psi\,\pi^\pm$ mass plot, 
choosing the higher of the two masses, and fit this distribution 
to study the new peak.  Results are shown in Fig. \ref{fig:Zcfit}.  
The fit includes an $S$-wave Breit-Wigner convolved with our 
Monte-Carlo-determined resolution on top of a four-parameter 
background.  We note that phase-space Monte-Carlo events, shown 
as a dashed line, have a shape very similar to this empirical background.  

The parameters of the resonant term extracted from this fit are: 
$$    M   = (3899.0 \pm  3.6 \pm  4.9) \;{\rm MeV}/c^2 \qquad
   \Gamma = (  46   \pm 10   \pm 20  ) \;{\rm MeV} \;.$$
We also determine the fraction of $J/\psi \pi^+\pi^-$ events 
that are in the $Z_c$ peak to be $(21.5 \pm 3.3 \pm 7.5)\%$.  

Similar results confirming our observations have been presented 
by Belle\cite{Z_c_Belle} and Northwestern University\cite{Z_c_NWU}.

\section{Selected Charmonium Results}

At $e^+e^-$ colliders, one can directly produce states with 
$J^{PC} = 1^{--}$.  
Specifically, these are the $^3S_1$ state of charmonium 
(in the usual $^{2S+1}{\cal L}_J$ spectroscopic notation), 
including for example the $J/\psi$ and $\psi(3686)$ (``$\psi'$'').  
Decays of these directly-produced states allow access to 
other charmonia such as the $\chi_{cJ}$, $h_c$, and $\eta_c$, 
which are the $^3P_{0,1,2}$, $^1P_1$, and $^1S_0$ states, respectively.  

In the following section, we give examples of some analyses 
involving these states, based on our first data samples from 2009.

\subsection{\boldmath A Limit on $J/\psi \to e\mu$}

At current sensitivities, lepton flavor violation is negligible 
in the Standard Model due to the very small neutrino masses.  
We perform a search for the flavor-violating decay 
$J/\psi \to e\mu$\cite{Jpsi_emu} in order to constrain models 
with new physics.    

This analysis is based on a sample of 225 million $J/\psi$ decays.  
We select two-track events with a back-to-back topology and 
veto on photons to remove radiative QED events.  
Electron identification requires a large value of the ratio of calorimeter 
energy to track momentum, $E/p$, with the muon detector used as a veto.  
Muon identification requires a small $E/p$ and a penetrating track 
in the muon detector.  
Our final signal variables are the total energy (calculated from 
the momenta and masses) and net three-momentum of the two detected particles.  

We find four candidates, with an expected background of $4.75 \pm 1.09$ 
determined from Monte-Carlo simulations.  
This yields a limit of: 
$${\cal B}(J/\psi \to e \mu) < 1.5 \times 10^{-7} \;.$$
This represents more than a factor of seven improvement over 
the best prior limit.  

Many other rare decays are accessible at BESIII, and we now have 
about five times more $J/\psi$ decays in our total data sample.

\subsection{\boldmath A Study of $\chi_{c0,2} \to \gamma\gamma$}

Our large sample of 106 million $\psi(3686)$ decays allows for 
precision studies of $\chi_{c2} \to \gamma\gamma$ decays\cite{chi_cJ}.  
We look for events with three photons and no tracks, arising 
from $\psi(3686) \to \gamma_1 \chi_{cJ}; \chi_{cJ} \to \gamma_2 \gamma_3$.  
We identify the $\chi_{cJ}$ states via the energy of the transition photon, 
$\gamma_1$, which has better resolution that the $\gamma_2\gamma_3$ 
invariant mass.  

\begin{figure}[htb]
\centering
\includegraphics[width=0.48\textwidth]{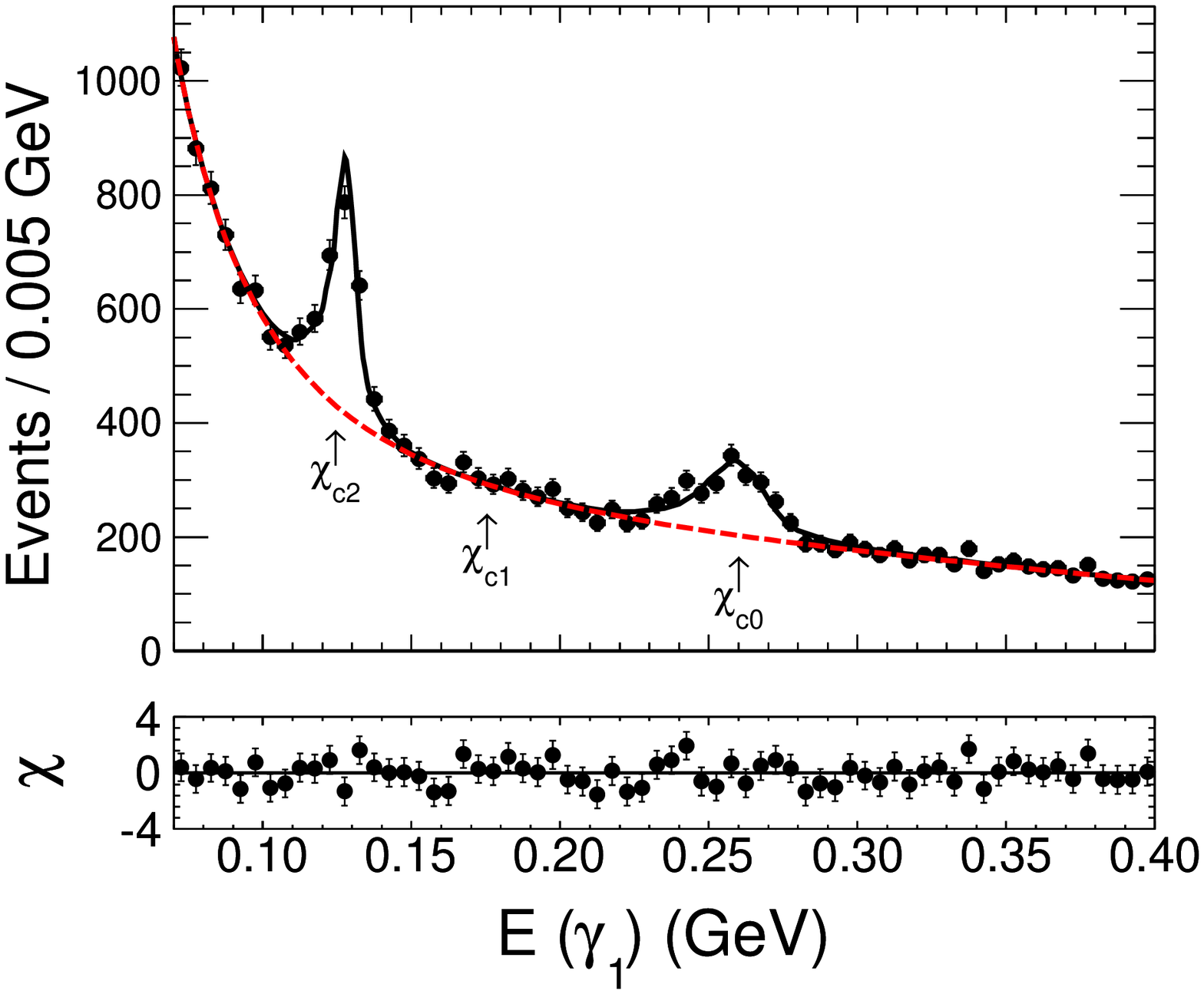}
\includegraphics[width=0.48\textwidth]{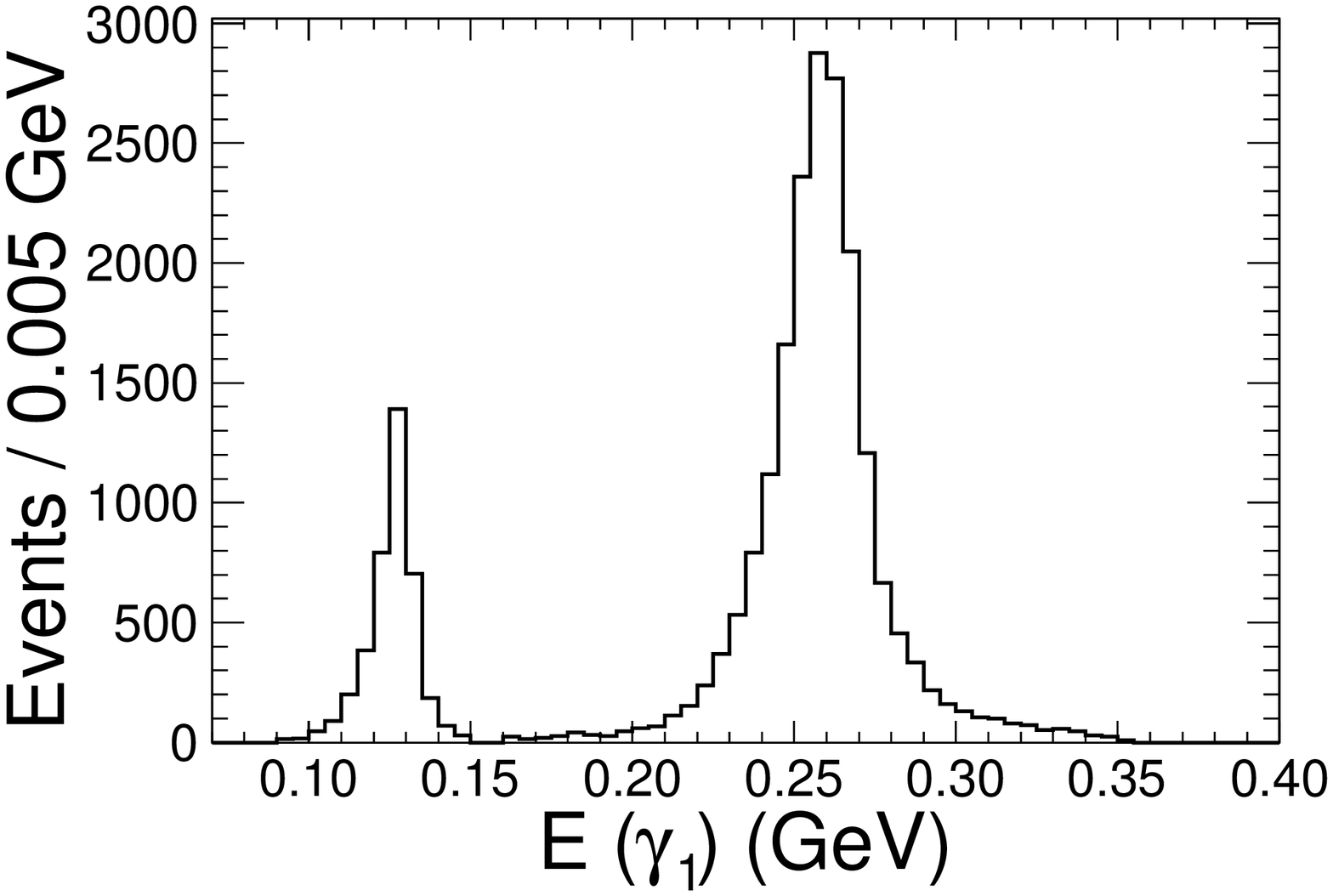}
\caption{Left: The $E_{\gamma_1}$ distribution, showing transition lines 
from the $\chi_{c2}, \chi_{c0}$.  
The data points are fit to a solid line which includes signal peaks 
above the red dashed background shape; fit residuals are displayed 
below.  
Right: Lineshape extracted from data, 
via $\psi' \to \gamma_1 \chi_{cJ}, \chi_{cJ} \to K^+K^-$.  }
\label{fig:chicJ}
\end{figure}

The transition lines are shown in Fig. \ref{fig:chicJ}; 
note that the $J=1$ transition is forbidden.  
Also displayed is the $E_{\gamma_1}$ lineshape we use, 
as extracted using a very high-purity sample (99.2\%) 
of hadronic decays of the $\chi_{cJ}$.  

We obtain the branching ratios: 
$${\cal B}(\chi_{c0} \to \gamma\gamma) 
  \,=\, (2.24 \pm 0.19 \pm 0.12 \pm 0.08) \times 10^{-4}$$ 
$${\cal B}(\chi_{c2} \to \gamma\gamma) 
  \,=\, (3.21 \pm 0.18 \pm 0.17 \pm 0.13) \times 10^{-4}$$
where the errors are from statistics, internal systematics, 
and PDG inputs\cite{PDG} on needed branching fractions and widths.  
We also extract the ratio of widths: 
$$ R \,=\, \Gamma_{2 \to \gamma\gamma}/\Gamma_{0 \to \gamma\gamma} 
    \,=\, (0.271 \pm 0.029 \pm 0.013 \pm 0.027) \;.$$  
The expected ratio is $4/15 \simeq 0.27$.  

Finally, we perform the first helicity analysis for the $J=2$ decay.  
We find that the ratio of helicity 0 to helicity 2 is 
$$f_{0/2} \,=\, 0.00 \pm 0.02 \pm 0.02$$ 
demonstrating the dominance of the helicity-2 process, as predicted 
by theory.

\subsection{\boldmath Masses and Widths of the $h_c$ and $\eta_c$}

This analyses uses the decay chain 
$\psi(3686) \to \pi^0 h_c, h_c \to \gamma \eta_c$ 
to study lineshapes of both the $h_c$ and $\eta_c$\cite{h_c}.  
The $\eta_c$ is reconstructed in sixteen exclusive channels. 
In fact, for five of the sixteen $\eta_c$ modes, we also report 
the first measurement of the branching fraction.  
The radiated $\pi^0$ is also detected, 
and this results in very clean peaks for both states, 
as shown in Figs. \ref{fig:etac} and \ref{fig:hc}.  

{
\begin{figure}[htb]
\centering
\includegraphics[width=0.48\textwidth]{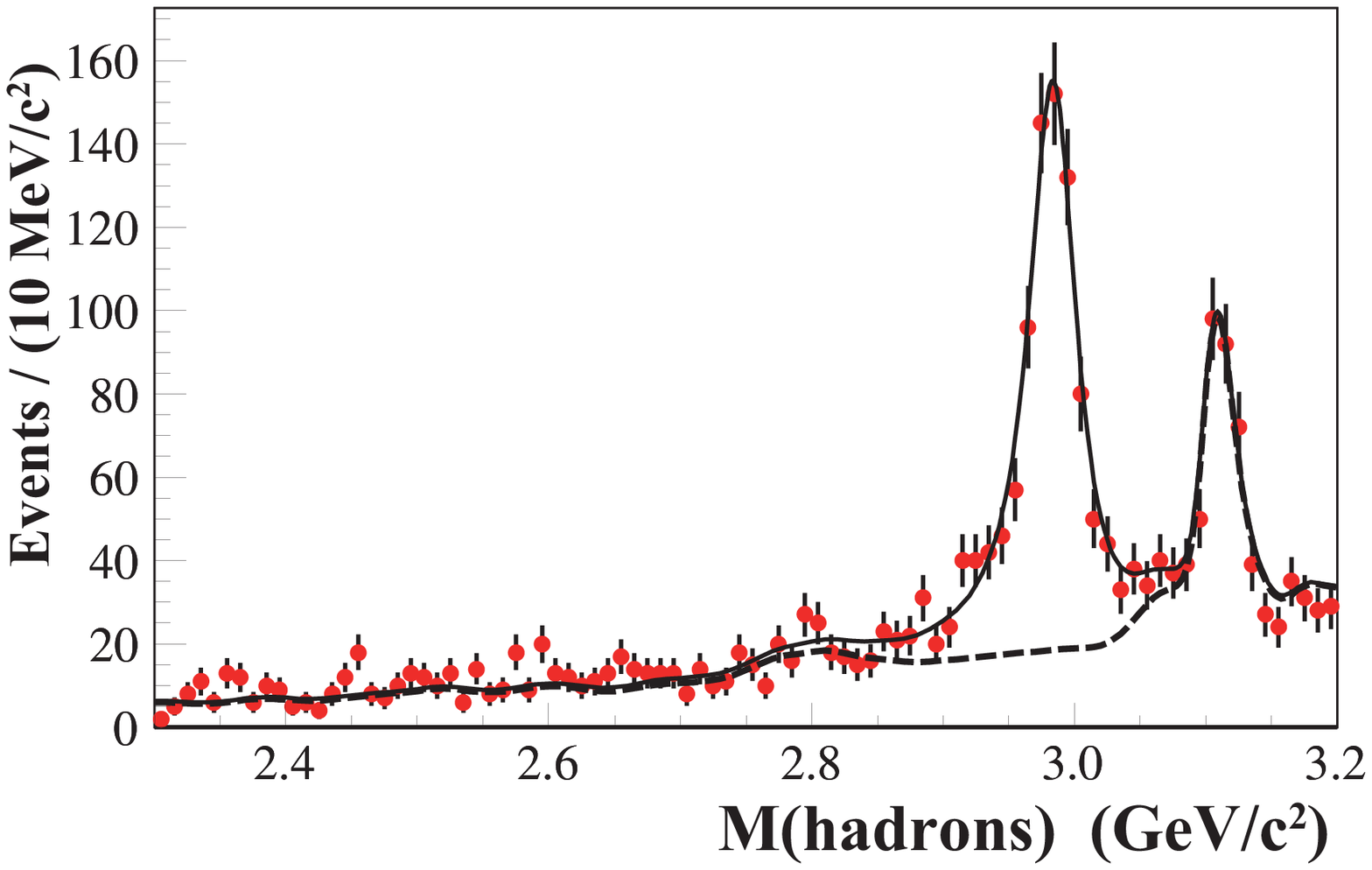}
\includegraphics[width=0.48\textwidth]{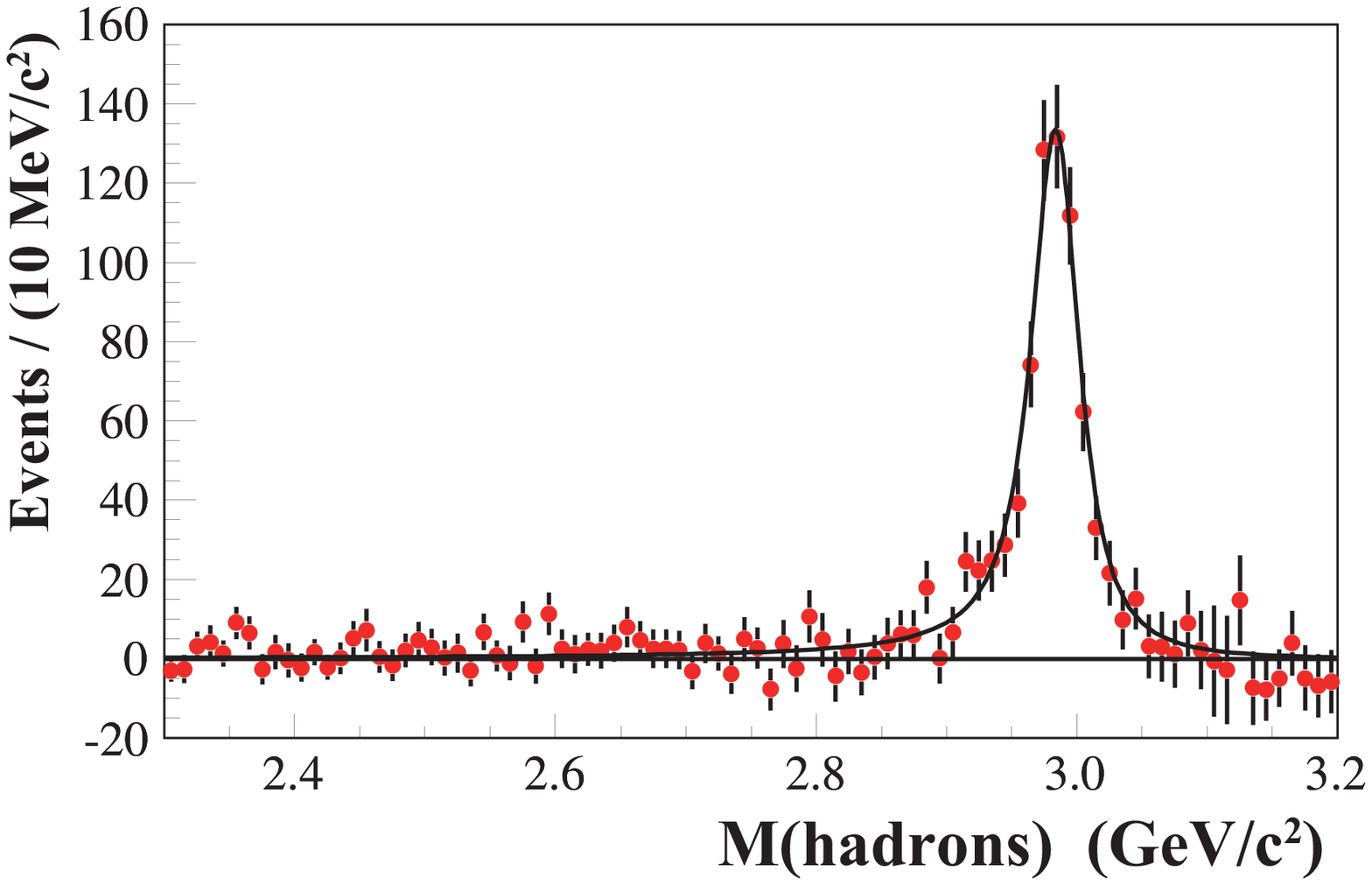}
\caption{Left: Sum of sixteen exclusive $\eta_c$ decay modes, 
with a fit including a peak from $\psi(3686)$ decays to 
the same final states and combinatorial background, 
in addition to the $\eta_c$ peak.  
Right: A background-subtracted plot of the same data.}
\label{fig:etac}
\end{figure}

\begin{figure}[htb]
\centering
\includegraphics[width=0.48\textwidth]{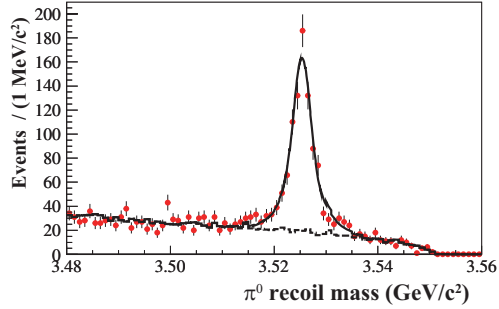}
\caption{The $h_c$ peak obtained via the $\pi^0$ recoil mass.}
\label{fig:hc}
\end{figure}
}

We obtain the most precise $h_c$ mass and width to date:  
$$     M(   h_c) = (3525.31 \pm 0.11 \pm 0.14) \;{\rm MeV}/c^2 \qquad
  \Gamma(   h_c) = (   0.70 \pm 0.28 \pm 0.22) \;{\rm MeV} \;.$$ 
The $\eta_c$ results, are also quite precise: 
$$     M(\eta_c) = (2984.49 \pm 1.16 \pm 0.52) \;{\rm MeV}/c^2 \qquad
  \Gamma(\eta_c) = (  36.4  \pm 3.2  \pm 1.7 ) \;{\rm MeV} \;.$$
While there are more precise width results available, the precision 
on the mass is similar to the best previous measurements\cite{PDG}.  
Those results have some tension with each other; ours favors a mass 
toward the higher end of their range.  
Furthermore, our $\eta_c$ measurements benefit from having highly 
suppressed interference effects due to our technique; 
this may be the best method with larger datasets in the future.

\subsection{\boldmath Mass and Width of the $\eta_c(2S)$}

\begin{figure}[htb]
\centering
\includegraphics[width=0.48\textwidth]{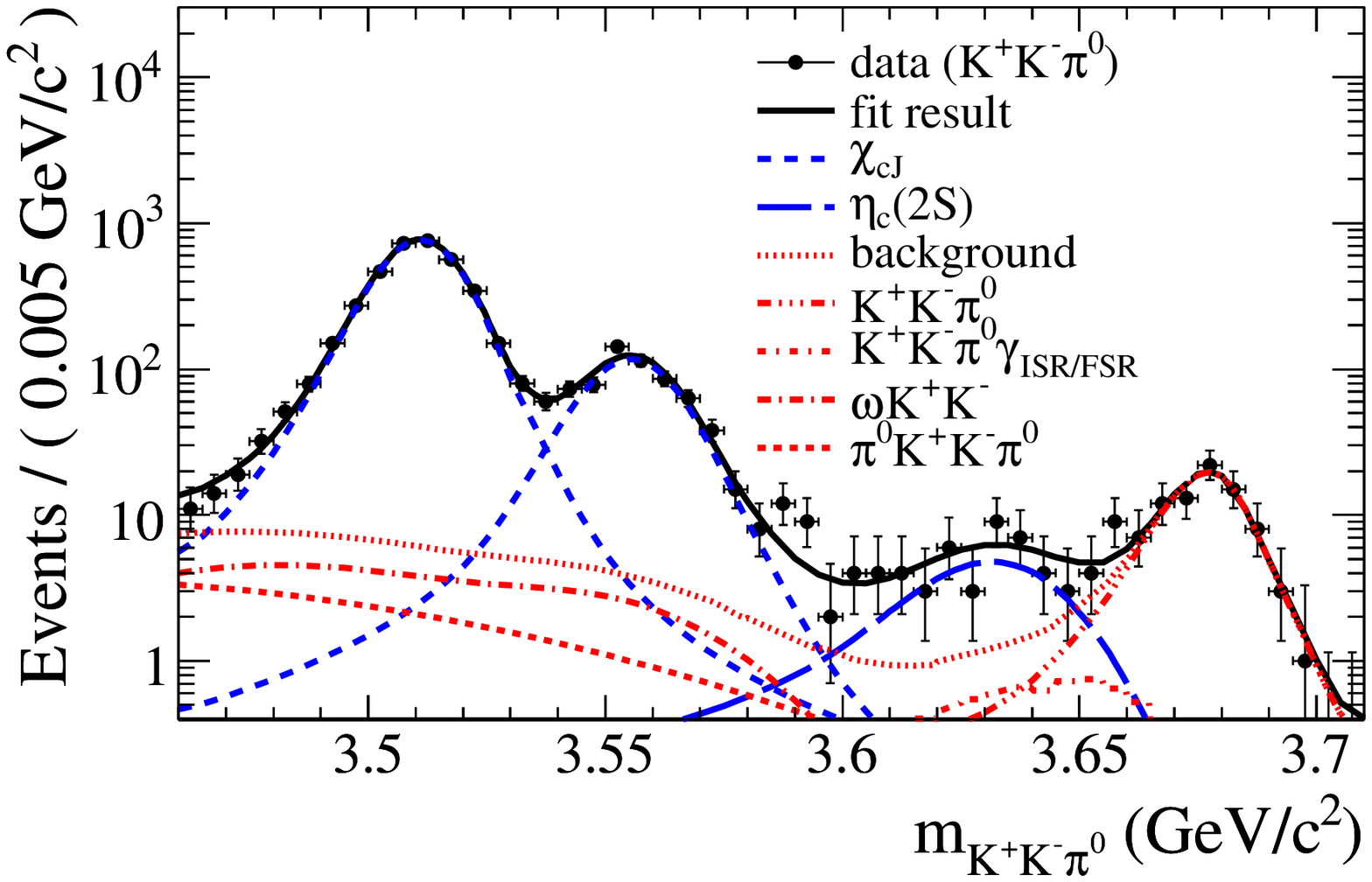}
\includegraphics[width=0.48\textwidth]{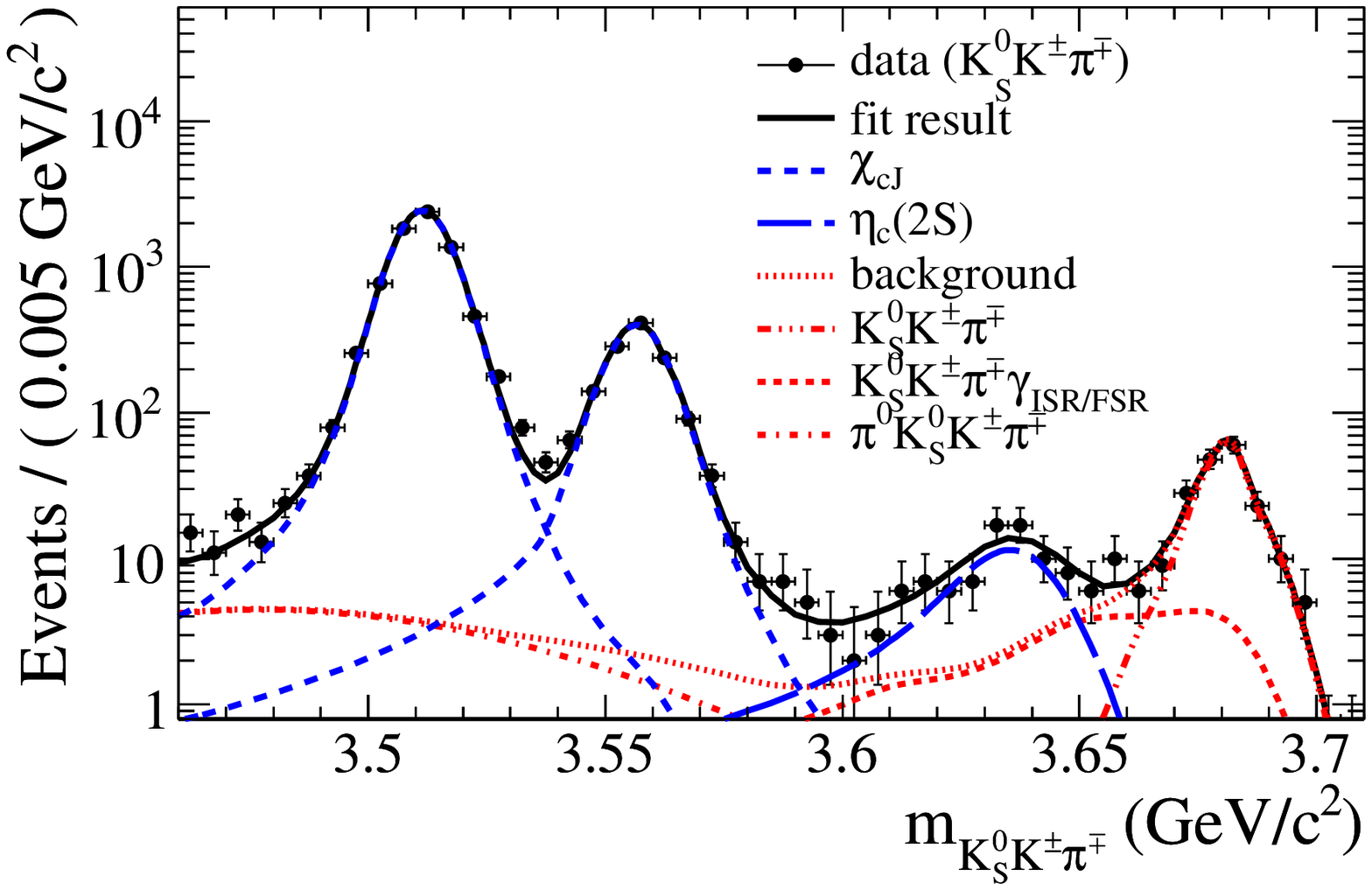}
\caption{Invariant mass of $K_S K^\pm \pi^\mp$ (left) 
and $K^+K^-\pi^0$ (right); the $\eta_c(2S)$ peaks are near 3.64 GeV.  
The many component curves are detailed in the caption. }
\label{fig:etac2S}
\end{figure}

BESIII has made the first observation of the $M1$ transition 
$\psi(3686) \to \gamma \eta_c(2S)$\cite{eta_c(2S)}.  
We reconstruct the $\eta_c(2S)$ in the $K_S K^\pm \pi^\mp, K^+K^-\pi^0$ 
modes and also detect the 48 MeV transition photon.  
A 4C kinematic fit enforcing four-momentum conservation is performed 
for both channels, and a 5C fit is also done for the mode with a 
$\pi^0$ where that particle's mass is the fifth constraint.  
As seen in Fig. \ref{fig:etac2S}, backgrounds from $\chi_{cJ}$ 
and $\psi(3686)$ decays are substantial, but a signal for the 
suppressed $M1$ transition is nonetheless evident.   

We extract the $\eta_c(2S)$ parameters: 
$$     M(\eta_c(2S)) = (3637.6 \pm 2.9 \pm 1.6) \;{\rm MeV}/c^2 \qquad
  \Gamma(\eta_c(2S)) = (  16.9 \pm 6.4 \pm 4.8) \;{\rm MeV} \;.$$
These are comparable in precision to the current PDG world averages\cite{PDG} 
of $(3637 \pm 4)$ MeV/$c^2$ and $(14 \pm 7)$ MeV, respectively.

\section{Preliminary Precision Charm Results}

BESIII has a sample of 2.9 fb$^{-1}$ of $\psi(3770)$ data; this 
resonance dominantly decays to $D^0\bar{D}^0$ and $D^+D^-$ pairs.  
At the peak, the total $D\bar{D}$ cross-section is about 6.6 nb.  
We note that there is insufficient energy for any additional 
hadrons; in particular, $D\bar{D}\pi$ is kinematically forbidden.  

A key to many analyses is the use of ``$D$ tagging''; 
this refers to full reconstruction of one $D$ or $\bar{D}$ meson 
in a fully hadronic final state.  Examples of modes used include 
$D^0 \to K^- \pi^+$ and $D^+ \to K^- \pi^+ \pi^+$.  

There are two key variables characterizing tags.  
The beam-constrained mass, $m_{bc} \,=\, \sqrt{E_{bm}^2 - p_{cand}^2}$, 
is based on momentum conservation.  Here, $E_{bm}$ is the beam 
energy and $p_{cand}$ is the $D$ candidate momentum (summed from 
the decay daughters).  
The energy difference, $\Delta E \,=\, E_{cand} - E_{bm}$, 
tests for energy conservation; here, $E_{cand}$ is the $D$ candidate 
energy.  Unlike $m_{bc}$, it depends on particle identification since charged 
daughter rest-masses are needed to calculate energies from track momenta.  

Use of tagging provides many advantages.  
First, it removes most backgrounds from continuum 
($u\bar{u}, d\bar{d}, s\bar{s}$ light-quark pair) events.  
Second, it constrains the kinematics of the other $D$.  
Specifically, it gives the vector direction of the momentum; 
the magnitude is known a-priori from energy-momentum conservation.  
And this constraint allows one to infer the four-vector of an 
unobserved neutrino in the decay of the $D$ opposite the tag 
as long as all other decay products are detected.  It is the 
last feature that is key to the two analyses discussed below.  

The branching ratios and efficiencies of the hadronic 
$D$ ($\bar{D}$) tagging modes largely cancel when studying 
the other ``signal'' $\bar{D}$ ($D$), since we measure the ratio 
of the tag plus signal yield to the tag only yield.  

In addition to the results presented here, 
work is in progress on many other topics.  
These include the strong $K\pi$ phase, 
quantum coherence measurements of modes including $K_S \pi^+ \pi^-$, 
the $D^0\bar{D}^0$ oscillation  parameter $y$, 
rate asymmetries in $D \to K_{L,S} n\pi$, 
the non-$D\bar{D}$ cross-section at the $\psi(3770)$, and more.

\subsection{\boldmath The Decay Constant $f_D$}

We now present BESIII's precise determination of the pseudoscalar 
decay constant $f_D$ from the decay $D^+ \to \mu \nu$\cite{munu}.  
This Cabibbo-suppressed $D^+$ mode has only been measured 
at $D\bar{D}$ threshold.  
On the other hand, $D_s^+ \to \mu\nu$ has been measured both at threshold 
and at $B$ factory energies; 
a submitted Belle result\cite{Ds_Belle} is currently the world's best. 

The decay rate is proportional $f_D^2$, which may be thought of 
as characterizing the probability that the $c$ and $\bar{d}$ 
quark overlap such that they may annihilate into a virtual $W^+$ boson.  
Other necessary external inputs include $V_{cd}$ and $\tau_{D^+}$; 
in particular:
$$ \Gamma(D^+ \to \mu\nu) \;=\; \frac{G_F^2}{8\pi} \, f_D^2 \, |V_{cd}^2| 
     \, m_\mu^2 \, m_{D^+}^2 \left( 1 - \frac{m_\mu^2}{m_D^2} \right) \;.$$

\begin{figure}[htb]
\centering
\includegraphics[width=0.9\textwidth]{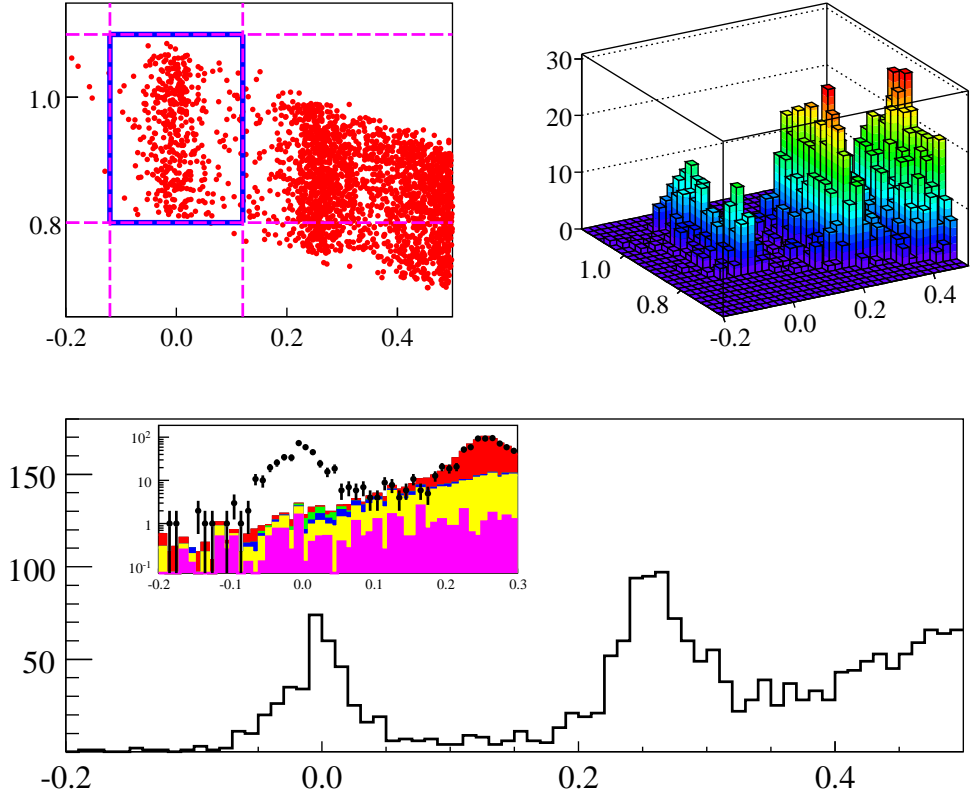}
\caption{Upper left: $\mu$ momentum, $p_\mu$ (GeV), vs. missing-mass-squared, 
$MM^2$ (GeV$^2/c^4$).  Upper right: lego plot of $p_\mu$ vs. $MM^2$.  
Bottom: $MM^2$ after the indicated $p_\mu$ cut.  
Bottom inset: log plots of $MM^2$ with stacked backgrounds in color 
(non-$D\bar{D}$ events in magenta, other $D\bar{D}$ in yellow, 
$D^+ \to \tau^+ \nu$ in blue, $D^+ \to \pi^+\pi^0$ in green, 
and $D+ \to K_L \pi^+$ in red ).}
\label{fig:munu}
\end{figure}

$D^+D^-$ events are tagged with nine $D^+$ decay modes: 
$K^-\pi^+\pi^+$, $K^-\pi^+\pi^+\pi^0$, $K^-\pi^+\pi^+\pi^+\pi^-$, 
$K_S\pi^+$, $K_S\pi^+\pi^0$, $K_S\pi^+\pi^+\pi^- $, 
$K_SK^+$, $K^+K^-\pi^+$, and $\pi^+\pi^+\pi^-$.  
We require exactly one track in addition to the tag, with the correct 
charge.  We veto any unused high-energy ($> 300$ MeV) EM calorimeter showers 
not matched to a track.  
This is especially effective in reducing $D^+ \to \pi^+ \pi^0$ background, 
which is important since $m_\pi^2$ is comparable to our $MM^2$ resolution.  

Our final signal variable is $MM^2 = E_{miss}^2 - p_{miss}^2$, 
where the missing four-momentum is obtained by subtracting the $D$ tag 
and signal $\mu$ momenta from the known initial-state four-vector.  
This quantity will peak at zero when only a neutrino was 
omitted from our kinematic calculations.  
Our data is presented in  Fig. \ref{fig:munu}, where a remarkably 
clean signal peak is evident.    

We observe a signal of $377.3 \pm 20.6 \pm 2.6$ events above a background 
of 47.7 events.  From this signal, we extract:
$${\cal B}(D^+ \to \mu \nu) \,=\, (3.74 \pm 0.21 \pm 0.06) \times 10^{-4}$$
$$f_D \,=\, (203.01 \pm 5.72 \pm 1.97) \;{\rm  MeV} \;.$$ 
This is more precise than the previous best measurement of 
$f_D \,=\, (205.8 \pm 8.5 \pm 2.5)$ MeV, based on 
818 pb$^{-1}$ from CLEO-c\cite{fD_CLEO}.  
It is in agreement with recent lattice QCD calculations; see, for example, 
the summary in Ref. \cite{munu}.  
Note that the measurement is still statistics-limited; 
we expect that BESIII will take more data in the future in order 
to further improve this important result.

\subsection{\boldmath $D$ Semileptonic Form-Factors}

BESIII has extracted the form-factors $f_{\pi,K}(q^2)$ 
from the semileptonic decays $D^0 \to K^- e^+ \nu, \pi^- e^+ \nu$\cite{henu}.  
Here, $q^2 = m_{e\nu}^2$ and these form factors describe the effects of 
meson structure in the decay, relative to idealized free-quark decay.  
In particular, the partial decay rate for $D^0 \to \pi^- e^+ \nu$is given by: 
$$  \frac{d\Gamma}{dq^2} \;=\; 
    \frac{G_F^2}{24\pi^3} \, |V_{ud}|^2 \, p_\pi^3 \, |f_{\pi}(q^2)|^2 $$
and a similar expression for  $D^0 \to K^- e^+ \nu$.  

The four tag modes $K^- \pi^+$, $K^- \pi^+\pi^0$, $K^- \pi^+\pi^+\pi^-$ 
and $K^- \pi^+\pi^0\pi^0$ are used.  Particle identification is important 
for both the $D$ tag and the semileptonic ``signal'' $D$.  
However, there is actually excellent {\it kinematic} separation 
between the Cabibbo-allowed $K e \nu$ mode and the ten-times rarer 
Cabibbo-suppressed $\pi e \nu$ mode.  This is in marked contrast 
to older analyses based on $D^*$ tagging with higher-energy $D$ mesons.  
With the very large luminosities of $B$ factories, it is now possible 
to use a full-event reconstruction tagging technique, which is 
far superior to the older $D^*$ tagging, but still has higher backgrounds 
than analyses from charm threshold.  

We require exactly two oppositely-charged tracks in addition to our 
hadronic $D$ tag, with the correct electron charge. 
Electron identification is based on $E/p$, 
while $K - \pi$ separation employs time-of-flight and $dE/dx$.  
We veto any unused high-energy ($> 250$ MeV) EM calorimeter showers 
not matched to a track.  
Our final signal variable is $U = E_{miss} - p_{miss}$; 
the ``miss''  quantities, representing the unobserved neutrino, 
are analogous to those in the previous analysis.  
For signal, $U$ peaks at zero and is similar to a missing-mass-squared.  
Fits to the $U$ distributions in Fig. \ref{fig:henuU} 
lead to the branching fraction results shown in Table \ref{tab:henuBF}.  

{
\begin{figure}[htb]
\centering
\includegraphics[width=0.48\textwidth]{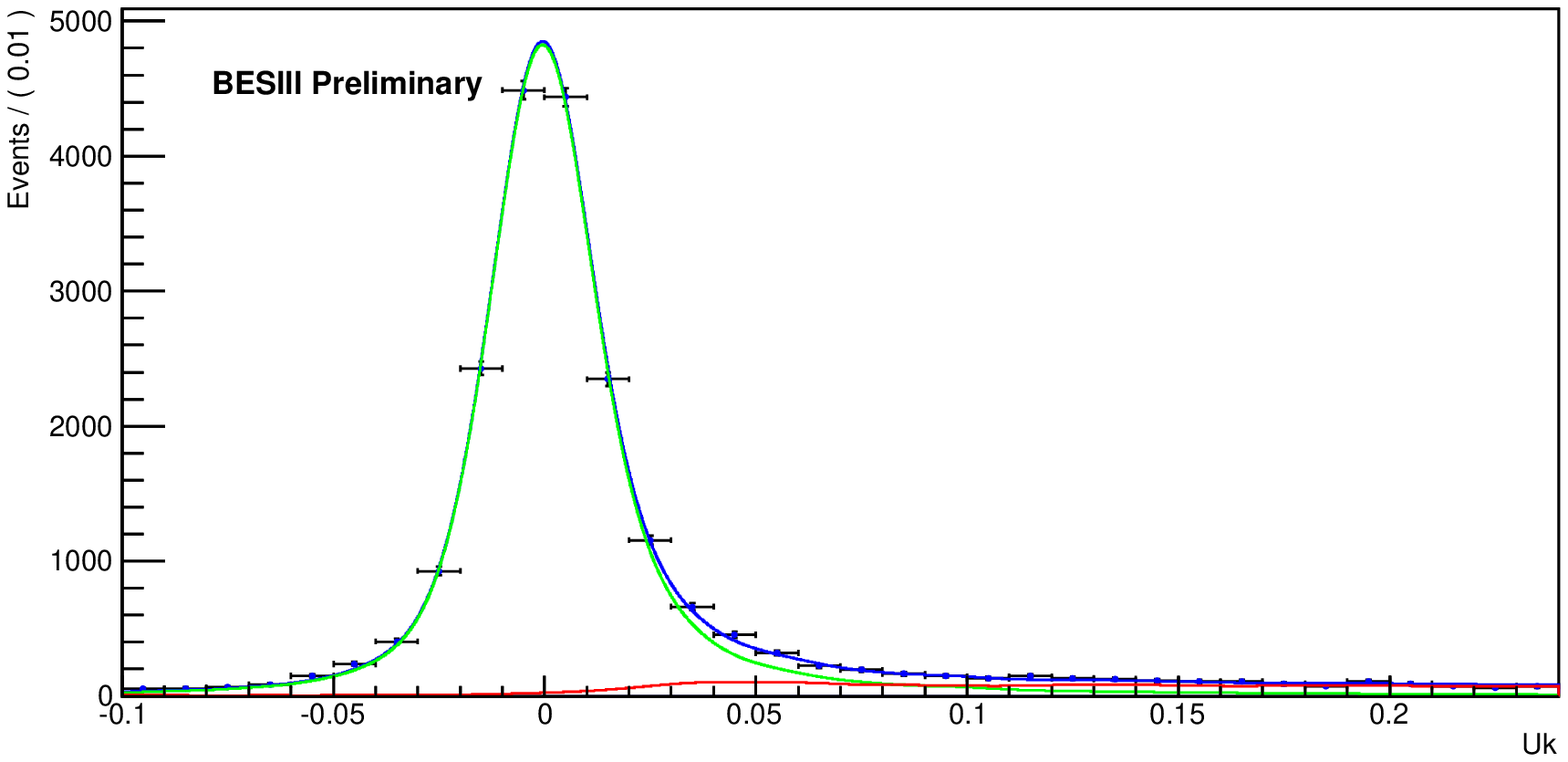}
\includegraphics[width=0.48\textwidth]{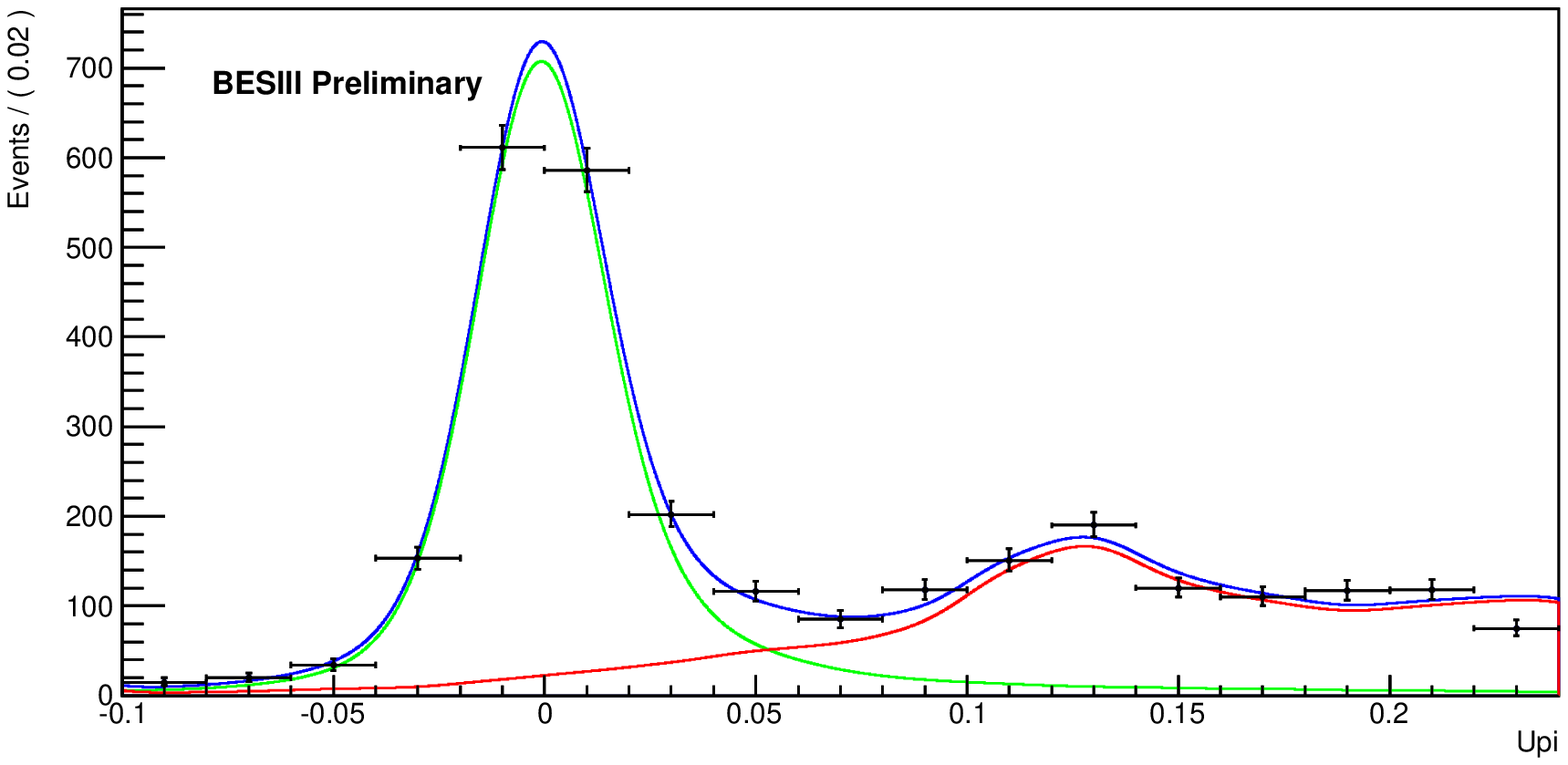}
\caption{$U \,=\, E_{miss} - p_{miss}$ distributions (GeV) for 
$D^0 \to K^-e^+\nu$ (left), $D^0 \to \pi^-e^+\nu$ (right).  
The blue total fit curve is the sum of a green signal shape and 
a red background term.}
\label{fig:henuU}
\end{figure}

\begin{figure}[htb]
\centering
\includegraphics[width=0.40\textwidth]{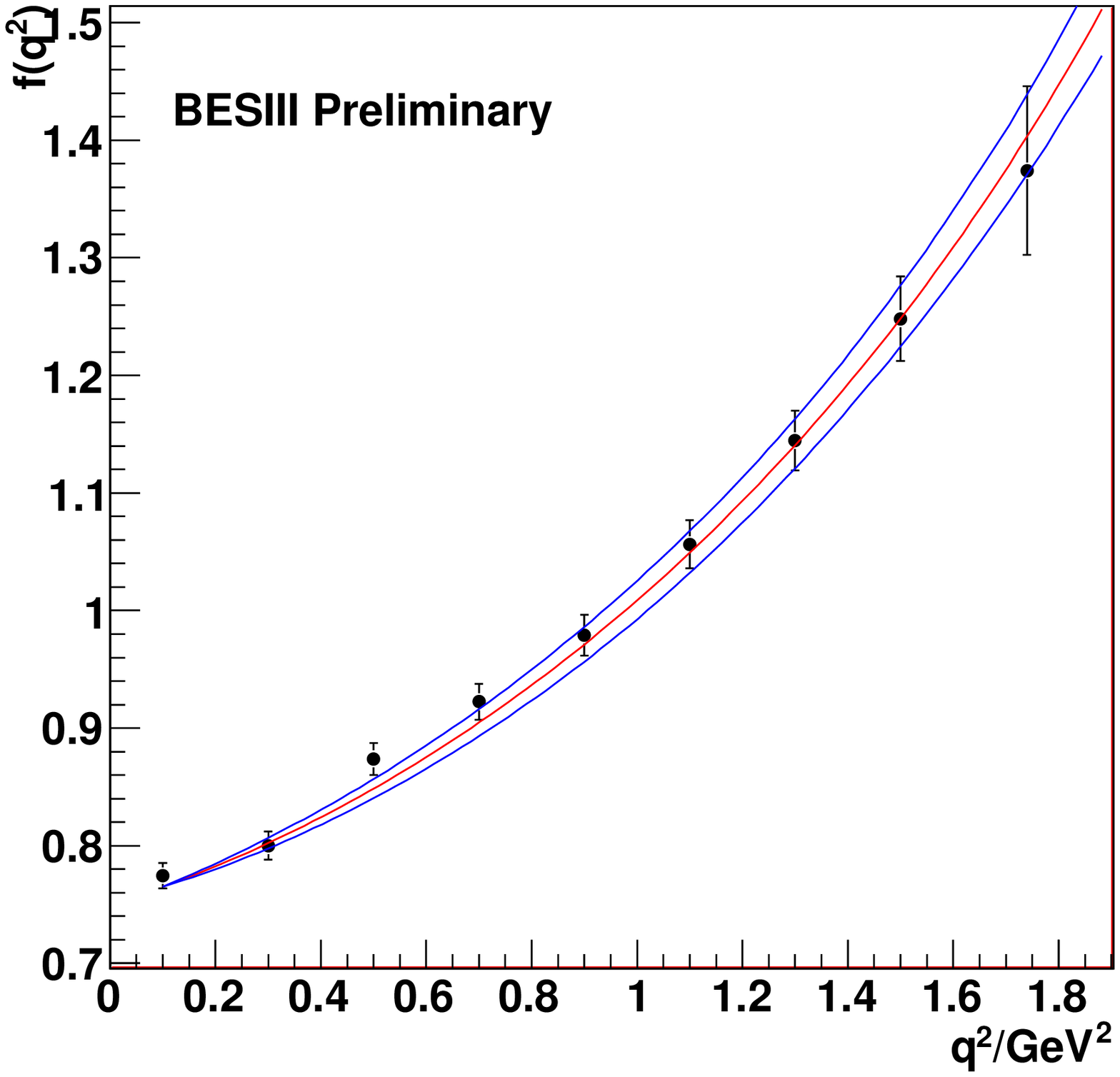}
\includegraphics[width=0.40\textwidth]{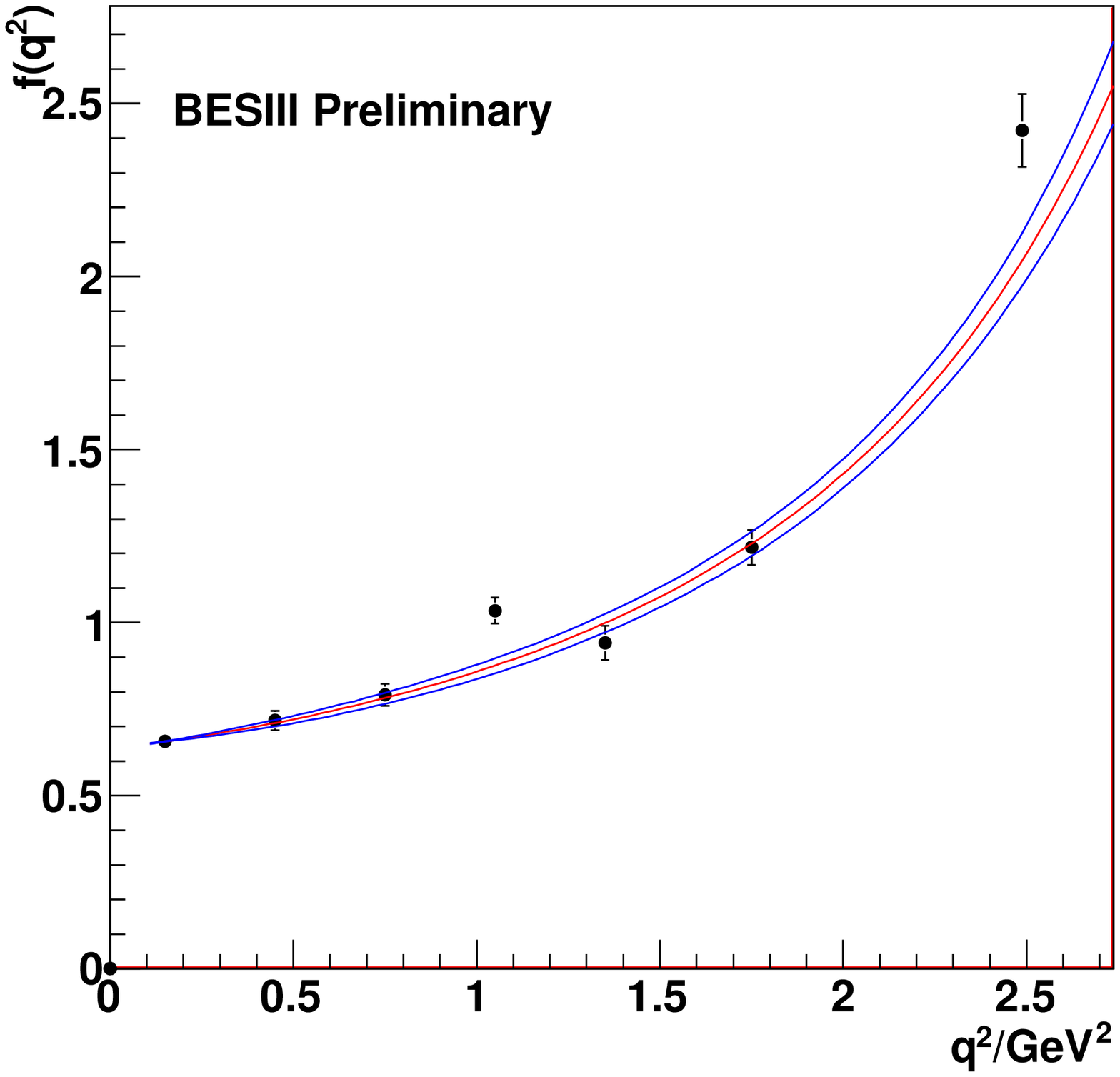}
\caption{Extracted form factors, $f(q^2)$, 
for $D^0 \to K^-e^+\nu$ (left), $D^0 \to \pi^-e^+\nu$ (right).  
The data points are compared to Lattice predictions (red) bracketed 
by one-sigma error bands (blue).  }
\label{fig:henuFF}
\end{figure}

\begin{table}[b]
\begin{center}
\caption{Preliminary BESIII branching fractions 
         and PDG 2012 world averages\cite{PDG}.}
\begin{tabular}{|l|cc|}  
\hline
Mode & BESIII BF (\%) & PDG BF (\%) \\
\hline
$D^0 \to   K^-e^+\nu$ & $3.542 \pm 0.030 \pm 0.067$ & $3.55  \pm 0.04$  \\
$D^0 \to \pi^-e^+\nu$ & $0.288 \pm 0.008 \pm 0.005$ & $0.289 \pm 0.008$ \\
\hline
\end{tabular}
\label{tab:henuBF}
\end{center}
\end{table}

\begin{table}[b]
\begin{center}
\caption{Preliminary BESIII form-factor results.  
For brevity, only results from the three-parameter series fit are shown.}
\begin{tabular}{|l|ccc|}  
\hline
Mode & $f_+(0)|V_{cs(d)}|$ & $r_1$ & $r_2$ \\
\hline
$D^0 \to   K^-e^+\nu$ & $0.729 \pm 0.008 \pm 0.007$  
          & $-2.179 \pm 0.355 \pm 0.053$ & $4.359 \pm 8.927 \pm 1.103$ \\
$D^0 \to \pi^-e^+\nu$ & $0.144 \pm 0.005 \pm 0.002$
          & $-2.728 \pm 0.482 \pm 0.076$ & $4.194 \pm 3.122 \pm 0.448$ \\
\hline
\end{tabular}
\label{tab:henuFF}
\end{center}
\end{table}
}
\clearpage

For the form-factor analysis, we divide the data into bins of $q^2$ 
to determine values of $d \Gamma/dq^2$ integrated over these bins.  
We note that our $q^2$ resolution is excellent and the smearing effects, 
which we do include, are modest.  
The extracted form factors are shown in Fig. \ref{fig:henuFF}, 
along with a representative lattice QCD calculation\cite{LQCD}.  
We have not attempted to update the comparison to lattice QCD made 
at the original presentation of BESIII results at CHARM2012; 
in the future, we will compare final BESIII results 
to all updated LQCD results.     
Numerical values of our form factor fit results are given in 
Table \ref{tab:henuFF} for the three-parameter version of the popular series 
expansion\cite{BechHill} prescription for parameterizing the form factor.  
The results for other fits are available in Ref. \cite{henu}.  

All results except for the $Ke\nu$ branching fraction are still 
statistics-limited.  The present results were obtained with 
about one-third of the full 2.9 fb$^{-1}$ sample;  
an update to the full dataset is expected soon.

\section{Conclusions}

We have presented a selection of results broadly spanning charm physics, 
including the discovery a possible new exotic state, 
studies of several conventional charmonium states, 
and first results from a precision $D$ physics program.  

Now five years from our first collisions, BESIII has established a broad 
and successful program in charm physics.  
Recently, in 2012, even larger samples have been accumulated at the $J/\psi$ 
and $\psi(3686)$; total samples are now about 1.2 billion and 0.35 billion 
decays, respectively.  
Furthermore, our 2013 dataset includes more data near 4260 MeV, 
and also a large sample at the $Y(4360)$.  
This and future running will sustain a vibrant physics program 
for many more years to come.

\Acknowledgements
We thank our BEPCII colleagues for the excellent luminosity 
and our BESIII collaborators for their many efforts culminating 
in the physics results presented herein.

\end{document}

%% file: econfmacros.tex
\def\beq{\begin{equation}}
\def\eeq#1{\label{#1}\end{equation}}
\def\eeqn{\end{equation}}

\def\beqa{\begin{eqnarray}}
\def\eeqa#1{\label{#1}\end{eqnarray}}
\def\eeqan{\end{eqnarray}}

\let\bar=\overbar

\def\Dslash{\not{\hbox{\kern-4pt $D$}}}
\def\dslash{\not{\hbox{\kern-2pt $\del$}}}

\def\msb{{\bar{\ssstyle M \kern -1pt S}}}

